\documentclass[article,onecolumn]{aa}

\usepackage[T2A]{fontenc}
\usepackage{graphicx}

\def\bsao{{Bull. Spec. Astrophys. Obs.}}
\def\ab{{Astrophys. Bull., }}

\newcommand{\arep}{Astronomy Reports }
\newcommand{\alet}{Astronomy Letters }

\begin{document}

\title{Results of selected stellar spectroscopy programs  at the 6-m telescope of SAO RAS  performed  with the NES echelle spectrograph}

\author{\large V.G.~Klochkova,
V.E.~Panchuk and M.V.~Yushkin}

\institute{Special  Astrophysical Observatory, RAS, Nizhnij Arkhyz, 369167 Russia \\
   \email{Valentina.R11@yandex.ru}
}

\titlerunning{NES results}
\authorrunning{Klochkova et al}

\abstract
{Over the past two decades the echelle  spectrograph NES  of the 6-m
telescope of the Special Astrophysical Observatory of the Russian
Academy of Sciences was used to perform  high resolution spectroscopy
of far evolved  stars spanning a wide range of initial masses.
The studies cover a diversity of stars  with high
mass-loss rates during the preceding and current stages of
evolution. All these stars have extended atmospheres and
structured circumstellar envelopes produced by strong stellar
winds. We have studied luminous blue variables (LBVs) near the
Eddington limit; hot supergiants with B[e] phenomenon in spectra,
which are very likely intermediate-mass binary systems soon
after the fast mass exchange stage; a group of yellow hypergiants,
as well as an extensive sample of low-mass post-AGB supergiants. The
diverse nature of the types of these stars whose common feature is
the presence of a circumstellar envelope makes the spectroscopy of
such objects a comprehensive task. Such studies consist of many etapes,
which include not only determining the peculiarities  of their atmospheric
chemical composition and understanding the role of supergiants in the
enrichment of the interstellar medium with freshly synthesized
elements, but also the determination of the evolution status  of
the objects considered, as well as search for and analysis of
spectroscopic manifestations of kinematic processes in their
extended and unstable atmospheres and gas-dust envelopes. We
spectroscopically monitored selected objects to study in detail
the instability of the kinematic state of the atmospheres of the
stars considered. Studies of stars at neighboring evolutionary
stages have been recently initiated. This review reports briefly
the most significant observational obtained within the framework
of the programs in 1998--2021.}

\maketitle

\section{Introduction}

The authors' (VGK and VEP) experience in stellar spectroscopy
performed with the 6-meter telescope of the Special Astrophysical
Observatory of the Russian Academy of Sciences accumulated during
the first twenty years of the operation of the 6-m telescope
\citep{Panch1998} allowed us to identify the types of
observational programs that seemed productive when carried out on
a multiprogram telescope. These are primarily programs whose
preparation involves no serious work input on the part of
qualified service personnel. We therefore focused our efforts on
the development of instruments permanently attached to the Nasmyth
foci platforms of the 6-m telescope \citep{ESPAC, Zebra, Lynx, Crab}.
Second, our experience has shown that programs that require
observations to be performed at predetermined time are difficult
to reconcile with the multi-program status of the telescope.

Our preference is for survey programs combined with continuous
monitoring of selected objects. In this case observations need not
to be made at predetermined times, which, given the information
about the repeatability of clear nights \citep{Erokhin} and seeing at
the 6-m telescope \citep{PAfan}, increases the probability of
successful observations. Of the programs that meet such
requirements, of interest to us are those aimed at studying
objects in the upper part of the Hertzsprung-Russell (H--R)
diagram. Objects studied span a wide range of parameters,
requiring quite a variety of skills acquired by analyzing the
results of the programs performed at the end of
the 20th century~\citep{KPan1998}.

The matter lost via stellar wind  and envelope outflow (or
its ejecting) is usually pre-processed through nucleo\-synthesis
during the preceding stages of evolution. During the evolutionary
stages under study, mass loss occurs at a varying rate and often
with a low level of symmetry. The lower limit of the velocities of
relative motions in the atmospheres and envelopes of the objects
studied is \mbox{$1$--$3$~km\,s$^{-1}$}, and that is why since
$1998$ we prefer to use the NES spectrograph of the  6-m telescope
\citep{OptTech,NES}, which provides better accuracy of
measurements based on individual lines and groups of lines.

\section{Low-mass  supergiants at the post-AGB phase}

We started carrying out this program by studying the supergiant
population at high galactic latitudes \citep{FHGL, UUHer} with
Main Stellar Spectrograph \citep{MSS} and the first
high-resolution LYNX echelle spectrograph \citep{Lynx}. With the
introduction of the NES spectrograph into the practice of
observations, the program was extended to
the spectroscopy of supergiants with large infrared flux excess
and proved to be most efficient both in terms of results and
scientometric criteria. IR flux excess is the main criterion for
selecting candidate stars, which include various star types with
high mass-loss rates during the previous and current stages of
evolution. The vast majority of the objects studied are asymptotic
giant branch (hereafter AGB) stars and their nearest  descendants --
post-AGB stars that rapidly evolve toward the planetary nebula (PN) stage
and for this reason also referred to as protoplanetary nebulae (PPN).
This type of stars is of interest primarily for searching for evolutionary
changes in the chemical composition of stars that have passed the AGB stage
and the third dredge-up.

\begin{figure*}
\includegraphics[angle=0,width=0.55\textwidth]{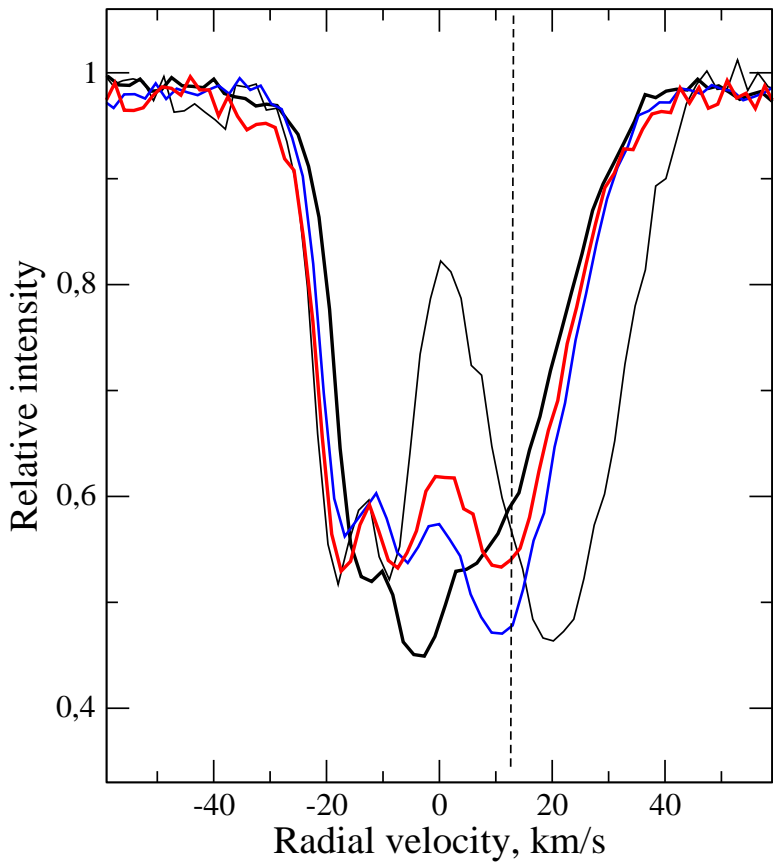}
\caption{The  Ba\,II\,$\lambda\,4934$~\AA\ absorption profile in the spectra of V5112\,Sgr
acquired on different dates: August $2, 2012$~(the thin solid line); June $13, 2011$~year
(the bold solid line); August $14,$ $2012$ (the red line), and July $14,$ $2001$ (the blue line).
The vertical dashed line shows the systemic velocity V$_{\rm sys}$=13~km/s and the envelope
components are indicated by arrows.}
\label{Ba4934}
\end{figure*}

Here we only point out the most significant results. First and
foremost, it is the detection of excess of $s$-process heavy
metals in the atmospheres of seven single post-AGB stars
\citep{rev1, rev2} in what was an empirical support for the
conclusions of the theory of the evolution of this type of stars.
In the spectra of three of these  stars a new
phenomenon---splitting of strong metal absorptions---was detected
for the first time, indicating that $s$-process heavy metals are
brought into circumstellar envelopes \citep{Enrich}.
Figure~\ref{Ba4934} excellently illustrates this effect showing
variable ${\rm Ba\,II}\,\lambda\,4934$~\AA\ absorption profile of
in the spectra of the post-AGB supergiant ${\rm V}5112\,{\rm
Sgr}$. According to data reported by~\citet{V5112Sgr}, two
short-wave components form in two different layers of the
structured envelope of the star, which expand with the velocities
of \mbox{$V_r=20$} and $30$~km\,s$^{-1}$ relative to the systemic
velocity.

The splitting of strong absorptions is observed in the optical
spectra of selected post-AGB stars with C-rich envelopes and
atmospheres enriched in carbon and heavy metals of the
$s$-process. The splitting is maximal in Ba\,II ions, whose lines
can break up into 2--3 components. An analysis of IR and radio
spectroscopy data showed that individual components of the split
absorptions form in structured circumstellar envelopes. We thus
observe the dredge-up of heavy metals synthesized during the
preceding evolution of the star into the shell. The type of strong
absorption profiles (splitting or asymmetry, number of components)
is associated with the morphology of the envelope and with its
kinematical and chemical properties.

The diversity nature of the types of infrared-excess stars makes
their spectroscopy a comprehensive task. The stages of their study
include not only identifying the features of the chemical
composition and elucidating the role of supergiants in the
enrichment of the interstellar medium with freshly synthesized
elements, but also the determination of the evolutionary status of
the objects considered and an  analysis of the spectral
manifestations of kinematic processes in their extended, often
unstable, atmospheres and in gas-dust envelopes. Studies of the
instability of spectra and of the kinematic state of atmospheres
of selected stars require spectroscopic monitoring.

The program that started with the detection of overabundance of
$s$-process elements has been extended to incorporate the
kinematics of stellar atmospheres and envelopes
\citep{HD56126,Envelopes}, the construction of spectral atlases
\citep{HD56126_atlas,Hypergiants_atlas} and spectropolarimetry of
selected objects \citep{Spolarimetry}. The commissioning of the
NES spectro\-graph made it possible to study circumstellar envelopes
by observing both the rotational structure of molecular spectra
(in emission and absorption, see Fig.\,7 in \citep{OptTech}) and
the narrow components of resonance lines in Fig.\,\ref{Ba4934}).

Long-term monitoring has revealed monotonic changes in the spectra
of several objects and allowed us to determine their evolutionary
status (see, e.g., \citep{IRC2, IRC3}). The status of some objects
remains uncertain as evidenced by the case of the peculiar
variable V838\,Mon that erupted in 2002.
\citet{V838Mon} determined the parameters and chemical composition
of its atmosphere from spectra acquired with the NES spectrograph
of the \mbox{6-m} telescope. However, the study of the nature of
the object and the cause of its outbreak continues for the third
decade with the results summarized by \citet{Kaminski}.

The NES spectrograph was used to determine the fundamental
parameters and detailed chemical composition of the atmospheres of
several rare types of stars.  For example, the spectra of
extremely hydrogen-deficient supergiants in common-envelope binary
systems were studied. Currently, information is available only for
four extremely hydrogen-deficient close binaries (HdBs) considered
to be SN\,Ia precursors. Surprisingly, all four objects have very
close effective temperatures $T_{\rm eff}\approx 10000$~K. The
best known among them is the supergiant $\upsilon$\,Sgr (${\rm Sp}
= {\rm A}\,2{\rm Ia}$), whose main features are strong and
variable H$\alpha$ emission and large IR excess. Based on
spectroscopic criteria only \citet{Kipper2012}
determined the effective temperature $T_{\rm eff}=12\,300$~K, surface
gravity $\log\,g=2.5$, and microturbulent velocity $\xi_t=9.5$~km\,s$^{-1}$
for this star. It has a hydrogen content of H/He\,$=3\times10^{-5}$,
small iron deficit [Fe/H]\,$=-0.8$ and altered abundances
of CNO-elements. A significant excess of heavy metals
[s/Fe]=+0.7 was also detected.  Many permitted and forbidden
emission features were found in the spectrum, which correspond to
low-excitation transitions of metal atoms and ions.  The
P\,Cyg-type profiles indicate the existence of a rotating
accretion disk in the system. \citet{Kipper2008} obtained similar
results  for a related system -- the semiregular variable KS\,Per.

\begin{figure*}
\includegraphics[angle=0,width=0.55\textwidth]{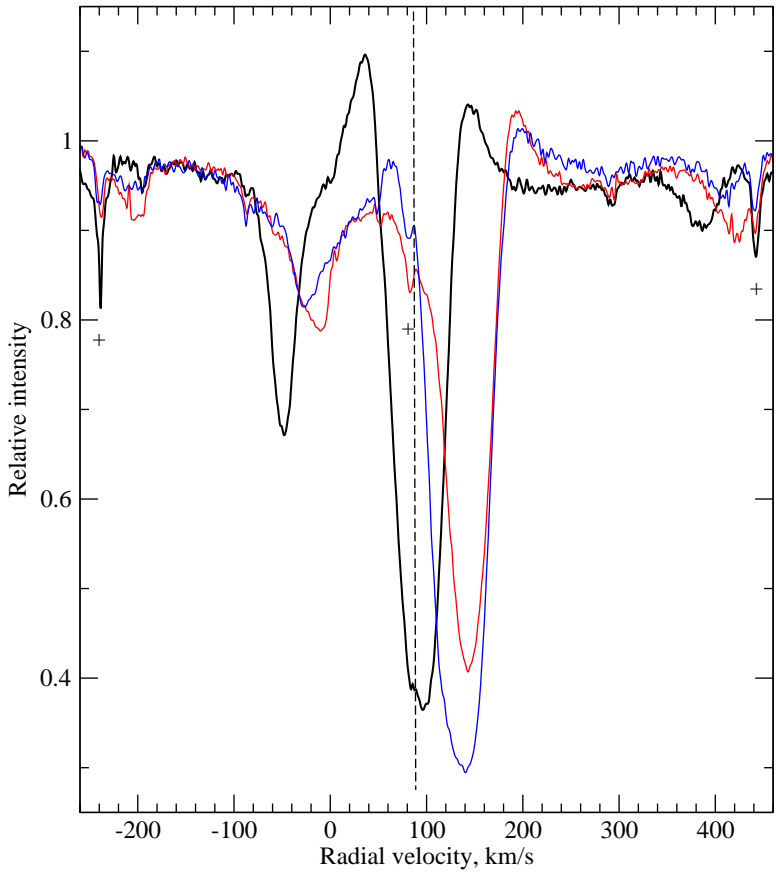}
\caption{Variable H$\alpha$ profile in the spectra of the yellow hypergiant
HD\,179821 acquired on different dates: September 24, 2010 (the blue line),
May 30, 2013 (the black line), and October 9,  2013 (the red line).
The vertical dashed line indicates the systemic velocity V$_{\rm sys}=86$~km\,s$^{-1}$
from \cite{Likkel}. The crosses indicate telluric absorptions.}
\label{HD179821}
\end{figure*}

One of the important advantages of the NES spectrograph is its
good sensitivity in the ground-based ultraviolet if combined with a
proper CCD.  This part of the spectrum is important for
spectroscopy of a number of specific programs. An example is the
spectroscopy of metal-deficient stars because the spectral
details required to analyze the features of the atmospheres of
such stars can be found just in the short-wave part of the
spectrum. The spectroscopic observations of 14 metal-deficient
G--K-type stars with high space velocities carried
out with the NES spectrograph in the wavelength range
3550--5100~\AA\ allowed the fundamental parameters and
atmospheric abundances of over $20$~chemical elements including
$s$- and $r$-process heavy elements ranging from Sr to Dy
\citep{SD} to be determined by the models atmospheres method.
For a number of elements, their abundances were computed taking
into account deviations from the local thermodynamic equilibrium
in the population of atomic levels. For six stars, the ratio of
the abundance of the long-lived radioactive element Th to that Eu,
which is an \mbox{$r$-process} element, was determined using the
method of synthetic spectrum. For the stars studied the various
Galactic populations to which they belong were determined based on
the inferred kinematical parameters and chemical composition.

An atlas~\citep{Atlas} of  high $S/N$ ratio spectra with high
spectral resolution was prepared based on observations made
in this poorly studied short-wavelength interval down to
$3055$~\AA~\citep{Atlas}. The spectra of well-studied stars with
similar temperatures ($\beta$\,Ori, $\alpha$\,Lyr and
$\alpha$\,Cyg) are compared with those of the metal-deficient
A-type supergiant KS\,Per with  hydrogen-deficient
atmosphere. The high spectral resolution and high $S/N$ ratio
make a detailed description of the spectra a relevant task
of  standard stars in the ground-based ultraviolet region
(300--380~nm), which is more informative than the
optical spectrum  due to line saturation, especially for hot
stars. Figure~\ref{HD179821} in \citep{Atlas} shows one of the shortest
wavelength fragments of this spectral atlas, $\Delta\lambda\,3100$--$3150$~\AA{}.

The spectra in the wavelength interval 3550--5000~\AA\ were also used
to produce an atlas for several metal-deficient stars, $-3.0<{\rm [Fe/H]}<-0.6$
\citep{ChJAA}. The above authors used these spectra to compute the
abundances of a large set of chemical elements in the atmospheres
of this sample of metal-deficient stars.

\section{Massive stars at  advanced  evolution stages}

The study of luminous stars is complicated by the fundamental
circumstance -- it is impossible to determine the star
center-of-mass velocity from the spectrum that forms in nonstationary
atmosphere. Whereas in the studies of post-AGB stars the results
of the millimeter- and submillimiter-wave spectroscopy of their
envelopes can be used to fix the systemic velocity, in the case of
massive supergiants the main circumstance contributing to the
determination of the star center-of-mass velocity is its
membership in a stellar group. A typical example is the study of
the spectrum of an LBV candidate in the  Cyg\,OB2
association \citep{No12CygOB2}. In such cases, it is necessary to
additionally measure the radial velocities of several faint member
stars of an association or open cluster in order to determine the
average radial velocity of the stellar group, and the task of
estimating the center-of-mass velocity of a luminous star becomes
rather time consuming \citep{CygOB2}.

The velocities of yellow hypergiants can be measured from CO, OH
molecular bands. For example, \citet{Oudm1996} reliably determined
the systemic velocity for the V1302\,Aql hypergiant
(the central star of the IR source  IRC+10420) from several
rotational bands of the CO molecule.  High-resolution optical
spectroscopy can be used to measure the positions of emissions
that form  in the circumstellar gas medium and thereby constrain
with high accuracy the velocity of the center of mass in the
``hypergiant+envelope'' system (e.g., for V1302\,Aql
\citep{IRC2} and for VES\,723 \citep{VES723}. The results
of long-term monitoring of Northern yellow hypergiants  are
summarized  and  their fundamental parameters summarized in
a review by \citet{rev2}. The fundamental parameters
of the stars were determined from homogeneous high-resolution
spectroscopic data. The luminosity criterion employed was the
infrared oxygen triplet O\,I $\lambda\,7773$~\AA\ with
extremal values of
equivalent width: the mean luminosity of the stars considered is
$\log L/L_{\odot}=5.43\pm0.14$. The combined data of
detailed positional measurements allowed the expansion velocity of
the circumstellar envelope to be estimated as
11--40~km/s. Weak absorptions allowed the
pulsation amplitudes of four objects to be estimated, which lie in
the narrow interval of $\Delta V_r=7$--$11$~km\,s$^{-1}$.

On the whole, the spectra of extremely luminous stars  compactly
located in the upper part of the Hertzsprung-Russell were shown to
exhibit a great variety of features, namely, the presence (or
absence) of permitted and forbidden emissions, emission components
of complex profiles, specific behavior  of spectral details of
different nature. High spectral resolution monitoring was shown to
be an efficient tool for detecting the variability of the dynamic
state at different depths in  the extended atmosphere and the
circumstellar envelope of hypergiants. In particular, such
observations made it possible to prove the reliability of the
yellow-hypergiant status for V1427\,Aql and the
absence of a companion in the  V509\,Cas system.

An general conclusion from observations of the  V1302\,Aql
hypergiant in $1992$--$2014$~years was that its effective
temperature has been increasing rapidly during the
20th~century---at a rate of about 120~K per year. Our
2001--2014 observations suggest that the hypergiant is
entering a phase where its temperature increase has ceased and the
object is approaching the Yellow Void boundary on the
Hertzsprung-Russell diagram.

Let us also mention the spectroscopy of several Cepheids
\citep{SUCas, Usenko}. The results for Polaris are particularly
important, the atmospheric parameters of the Cepheid member of
this nearby multiple system were studied in detail \citep{Usenko}.
Later, \citet{Polaris} were the first to determine  the basic
parameters of companion~B. \citet{Turner} used extensive data of
the spectral monitoring of Polaris carried out by V.G.~Klochkova and
M.V.~Yushkin with the NES spectrograph to study the variation of its
luminosity and effective temperature and concluded that Polaris pulsates
in the fundamental mode undergoes  the first crossing of the instability strip.

\section{CHANGES IN THE OPTICAL SPECTRUM OF THE $\rho$\,CAS HYPERGIANT DUE TO
A ENVELOPE EJECTION  IN 2013}

\begin{figure*}
\includegraphics[angle=0,width=0.65\textwidth]{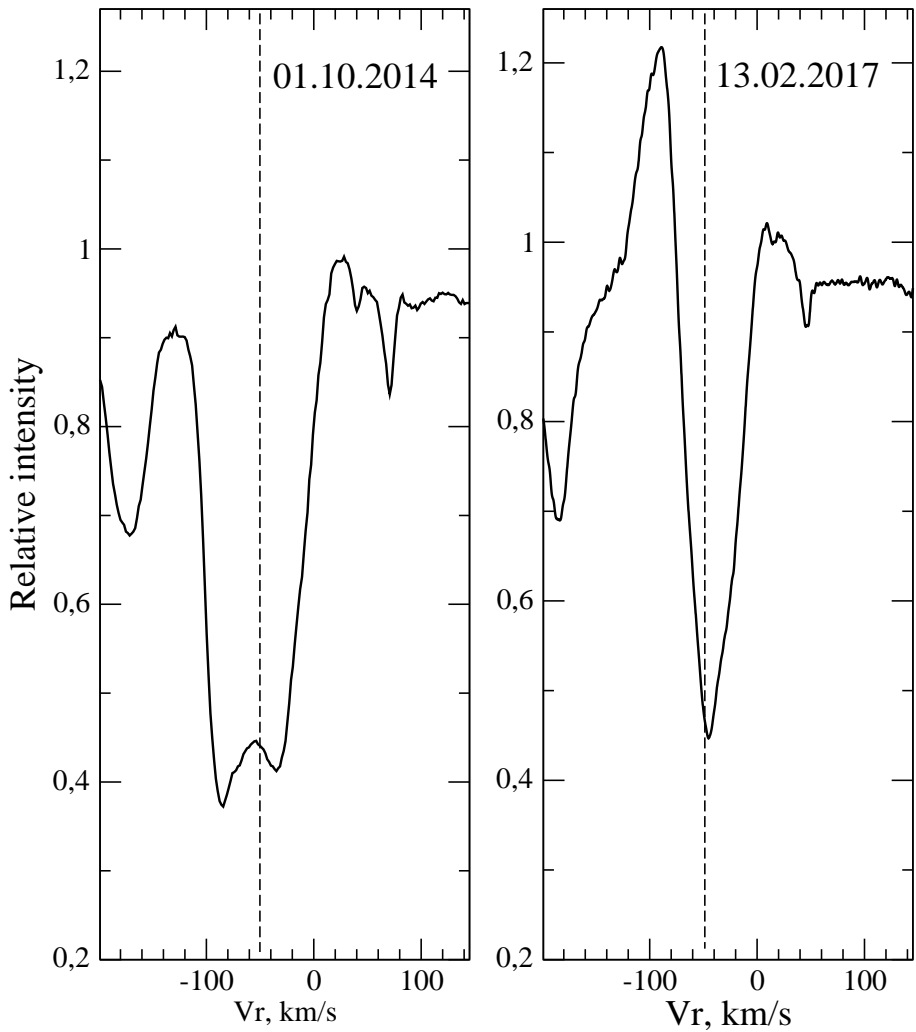}
\caption{H$\alpha$ profile in spectra $\rho$\,Cas: (a) after the $2013$~year outburst,
(b) after the 2017 brightness maximum.}
\label{rhoCas_Halpha}
\end{figure*}

Spectral monitoring of the yellow hypergiant $\rho$\,Cas with NES
spectrograph attached to the 6-m telescope after the mass ejection
that occurred  in 2013 revealed changes in H$\alpha$ line: from
a doubled core profile in $2014$ to the reverse
P\,Cyg-profile in the early $2017$ (see
Fig.\,\ref{rhoCas_Halpha}) and again to the doubled-core profile
with a strongly redshifted core, which is indicative of the fast
infall of matter. The splitting of the low-excitation absorption
profiles into three components \citep{rhoCas2018} was first
detected in $2017$. It was concluded that there is no correlation
in the evolution of the H$\alpha$ profiles and splitted absorption
profiles. Pulsation-type variability with an amplitude of about
10~km/s is characteristic solely of weak and
moderate-intensity absorption features. In the long-wave part of
the 2013 spectrum envelope emissions were detected whose
intensity in 2017 decreased until almost disappearing. Envelope
emission lines of metals are permanently present in the wings of
the H and K lines of Ca\,II. Monitoring of $\rho$\,Cas allowed us
to record dynamic instabilities in the star's upper atmosphere and
detect for the first time the stratification of its gas envelope
during the $2017$ mass-ejection episode. \citep{rhoCas2018}.

The $2013$ ejection was accompanied by a decrease of the
$\rho$\,Cas temperature and brightness by about $3000$~K and
$0\fm6$, respectively \citep{rhoCas2019}. During the ejection TiO
molecular bands and atmospheric low-excitation metal lines
characteristic of a later spectral type appeared in the spectrum.
An analysis of the parameters of emissions that appear in the
phases of maximum brightness showed that they vary synchronously
with the flux in the strong forbidden [Ca\,II] emission. Recent
ejections suggest that the time interval between such events is
decreasing. This could mean that $\rho$\,Cas is ``preparing'' for
a major mass ejection followed by a star crossing the Yellow Void
boundary. The relevance of systematic spectral monitoring of
$\rho$\,Cas is obvious.

\section{HOT SUPERGIANTS WITH B[E] PHENOMENON}

\begin{figure*}
\includegraphics[angle=0,width=0.8\textwidth]{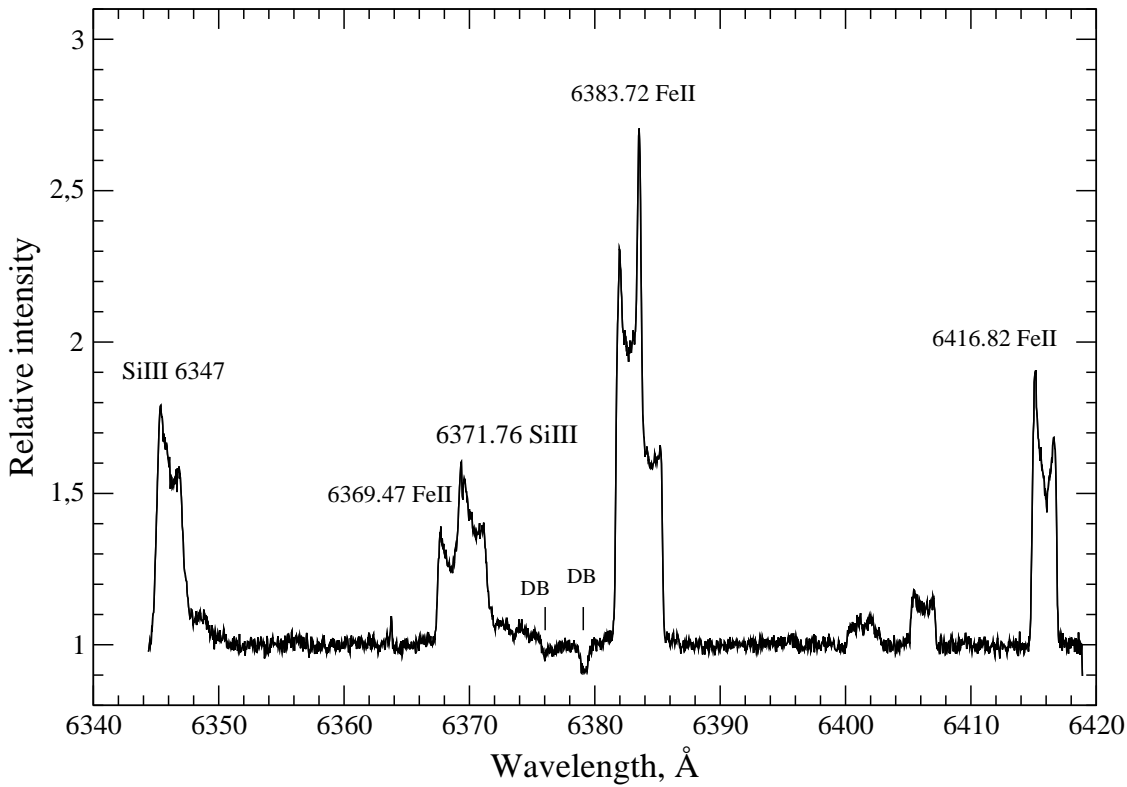}
\caption{Fragment of the spectrum of CI\,Cam acquired with the NES
spectrograph attached to the 6-m telescope on December 6, 2019.
Identification of the main features is shown.}
\label{CI_Cam}
\end{figure*}

As noted above, the main criterion for selecting candidates to the
family of luminous stars with infrared excess is the anomalous
spectral energy distribution. Due to such a broadly encompassing
criterion, the sample often included stars of unclear evolutionary
status. For example, before their detailed study began,
V1302\,Aql and 3\,Pup were considered to be  post-AGB stars evolving
toward the planetary nebula stage. Only a study of the chemical
composition of the atmospheres of V1302\,Aql \citep{IRC1} and
3\,Pup \citep{3Pup} allowed both stars to be classified as massive supergiants at
different stages of evolution. Currently, V1302\,Aql
is the most bona fide  representative of the yellow hypergiant
family, and the hot star 3\,Pup is classified as a B[e] object.

The B[e]-phenomenon consists in the spectrum of the star
exhibiting a number of peculiar details: strong H\,I and He\,I
emissions, as well as emissions of the permitted metal lines and
low-excitation forbidden lines. The second essential feature of
stars with the B[e]-phenomenon is their strong infrared flux
excess due to the presence of hot circumstellar dust. However, the
stars meeting the above two criteria constitute a group of very
heterogeneous objects. In the spectrum of 3\,Pup the
complex line profiles change with time \citep{3Pup_we}: the
magnitude and sign of the absorption asymmetry and the blue-to-red
emission intensity ratio vary. The profiles of all forbidden
emissions in the recorded wavelength range have the same shape
and width and the same radial velocity within the errors, which
allows us to adopt the average radial velocity of forbidden
emissions as the systemic velocity of the star: $V_{\rm sys}=28.5\pm0.5$~km/s.
Weak photospheric absorptions exhibit bona fide date-to-date velocity variations
(by up to 7~km/s.

According to its main parameters, 3\,Pup is close to the
star MWC\,17 with the phenomenon B[e] in its spectrum.
Multiple  observations performed with the NES spectrograph
attached to the 6-m BTA telescope made it possible to study in
details the features of the optical spectrum of MWC\,17
\citep{MWC17}. Numerous permitted and forbidden emissions, as well
as interstellar Na\,I lines and DIBs were identified in the
$4050$--$6750$~\AA\ wavelength interval. In this case too, the
velocity of forbidden emissions is adopted as the systemic
velocity $V_{\rm sys}$. However, a comparison of our data obtained
for MWC\,17 with earlier published data leads us to
conclude that in the spectrum of MWC\,17 shows no
significant variability unlike that of 3\,Pup.

\begin{figure*}
\includegraphics[angle=0,width=0.6\textwidth]{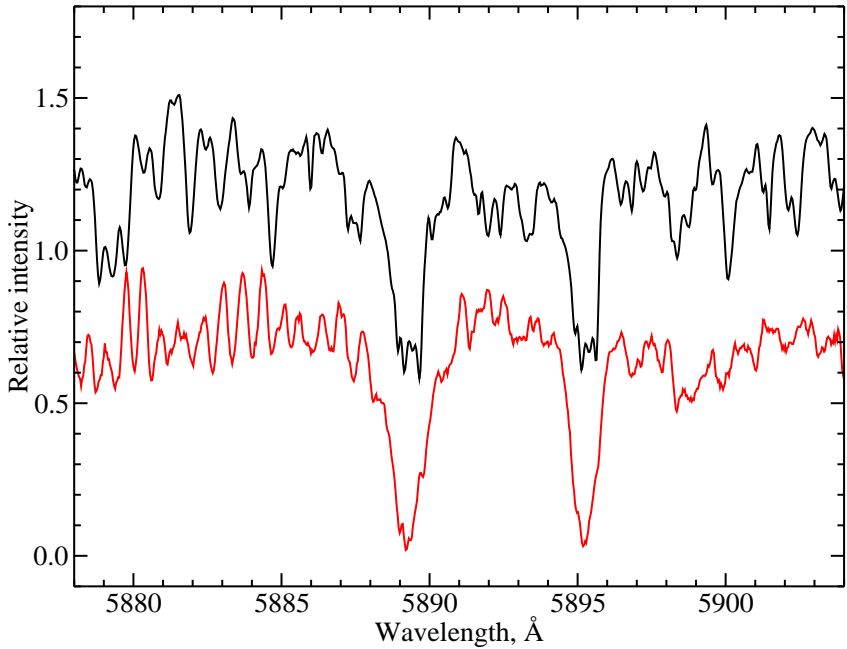}
\caption{Fragment of the spectrum of R\,Cam near the Na\,I resonance doublet
for the two light-curve phases. The upper spectrum is taken near the maximum, and
  the lower, near minimum visual brightness.}
    \label{RCam1}
\end{figure*}

Among the stars with the B[e]-phenomenon  B-supergiants in binary
systems have especially challenging spectra. As an example, we
mention the object MWC\,84\,=\,CI\,Cam, which was
observed spectroscopically over considerable time with the NES
spectrograph attached to the 6-m telescope. The object is unique
because it has both a companion and a circumstellar gas disk.
\citet{Bars} analyzed the results of the spectroscopic monitoring
of CI\,Cam and showed that the companion is a white dwarf.
\citet{Mir2002} found a complex structure of permitted and
forbidden emissions, which is indicative of the stratification of
their formation region. They also concluded that these
spectral features formed in a tilted circumstellar disk observed
edge-on. To illustrate this feature, we show a fragment of the
CI\,Cam spectrum in Fig.\,\ref{CI_Cam}. The Si\,III and Fe\,II ion
emissions are structured. Also apparent in this fragments are the
interstellar bands  DIB\,6376.32 and 6379.32\,\AA. We also
point out that all components of the infrared oxygen triplet
OI\,7773\,\AA\ in the spectrum of CI\,Cam appear as quite strong
and structured emissions.

\section{MIRA CETI TYPE STARS}

Miras are cool low-mass AGB stars (with masses less than
$1$--$3~M_{\odot}$. Spectroscopic observations performed
with the Main Stellar Spectrograph \citep{MSS} attached to the 6-m
telescope  in the late $1970$s, combined with computations of
synthetic \citep{TiO} molecule spectra suggested that the optical
spectra of Miras do not form in the stellar atmosphere, but mainly
in the cold circumstellar envelope \citep{Panch1978}. This
hypothesis can explain the variations of the visual amplitudes
from period to period, the radial-velocity amplitudes that do not
repeat from period to period, and the peculiarities of the
behavior of hydrogen and metal emission lines \citep{Mira_emis}.
The new program of spectroscopic observations of Mira stars
(performed with the NES spectrograph) mostly focuses on S-type
stars where the carbon-to-oxygen number ratio is close to unity.
In the atmospheres and envelopes of these objects almost all C and
O atoms are bound in CO molecules. The CO molecule has a maximum
dissociation potential, implying that near O/C\,$\approx1$
there is a lack of oxygen to form oxides and a lack of free carbon
to form carbon-containing molecules. Thus the atmospheres of
S-type Miras have significantly smaller content of TiO molecules
than those of M-type Miras, and exhibit virtually no ${\rm C}_2$
and CN molecules typical of C-type Miras. Therefore, the
atmospheres of S-type stars are the most transparent for this
interval of effective temperatures, allowing the dynamics of the
spectral features of the atmosphere and envelope to be observed
over a wide range of light-curve phases. Under certain conditions,
in the most opaque parts of the atmosphere the spectrum of the
circumstellar envelope can be observed. Figure\,\ref{RCam1} shows
the part of the spectrum in the vicinity of the Na\,I resonance
doublet for the two light-curve phases of the Mira variable star
R\,Cam  (its spectral type varies from
${\rm S}2.8{\rm e}$ to ${\rm S}8.7{\rm e}$).

\begin{figure*}
\includegraphics[angle=0,width=0.75\textwidth]{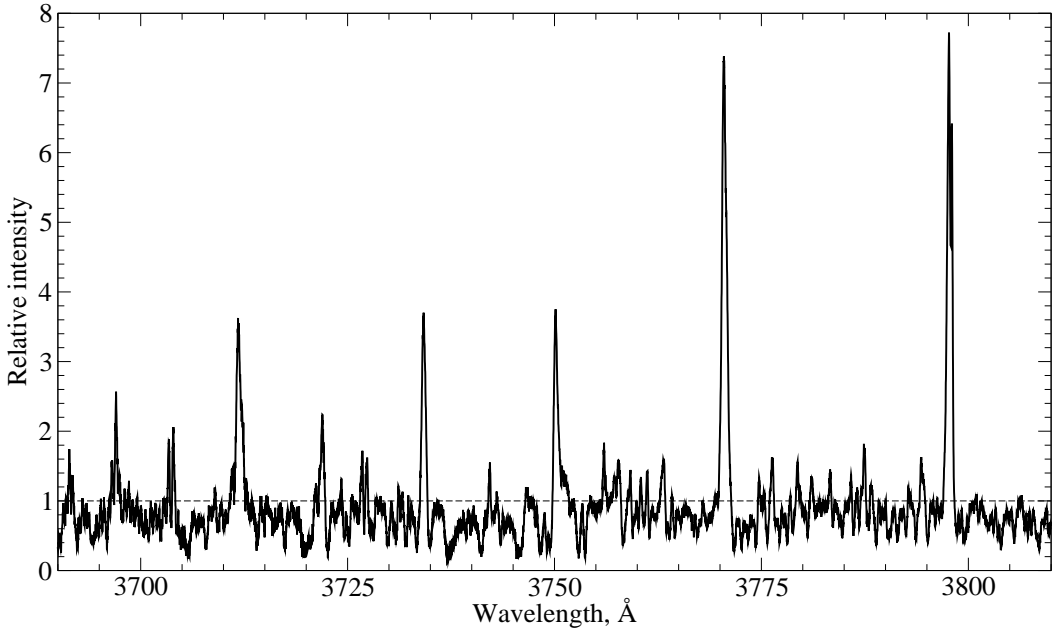}
\caption{Emissions of the Balmer lines (from H16\,3703.9~\AA, to H10\,3797.9~\AA)
and  numerous metal emissions and absorptions in a fragment of the spectrum of $\chi$\,Cyg -
  an S-type Mira.}
\label{RCam2}
\end{figure*}

\begin{figure*}
\includegraphics[angle=0,width=0.5\textwidth]{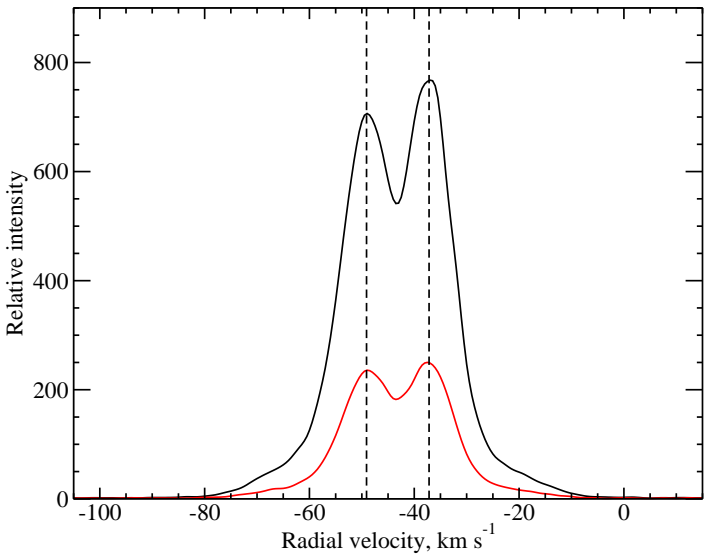}
\caption{Profiles of Bowen OIII lines in the Vy\,1$-$1 spectrum taken with the NES spectrograph.
The profile of the OIII\,$\lambda\,4559$~\AA\ line is marked in red, and that of the
OIII\,$\lambda$\,5007~\AA\ line, in black. Half the distance between the two dashed verticals
corresponds to the expansion velocity.}
\label{Vy-Oxy}
\end{figure*}

The high spectral resolution allows us to measure the widths of
the molecular lines (or their blends) in the
vibrational-rotational band $(2; 1)$ of the electronic system
$\gamma$' of the TiO molecule (the bottom spectrum in Fig.\,\ref{RCam1}, left part).
Near the maximum light narrow features repeating in the D$1$ and D$2$ lines (and hence certainly
belonging to sodium lines) were detected in the cores of the
Na\,I resonance doublet. The width of these features is smaller
than that of the lines of the molecular spectrum. All spectra have
good signal accumulation, implying than that the deep Na\,I
absorption cores are real. We believe that at maximum light, when
the molecular absorption does not distort the structure of the
Na\,I doublet cores, details of several envelopes detached from
the star during the AGB stage can be seen in the spectra of S-type
Miras. The structure of the cores of the resonant sodium doublet
lines in Miras was detected for the first time.

In the ground-based ultraviolet, where the molecular absorption is
insignificant, spectra of S-type Miras exhibit Balmer emission
lines with profiles indented by metal absorption lines
(Fig.\,\ref{RCam2}). The distortion of the emission lines can be
used to estimate the parameters of the absorbing layer located
above the formation region of emission features.

\begin{figure*}
\includegraphics[angle=0,width=0.6\textwidth]{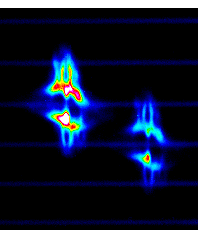}
\caption{Doppler image of a pair of  O\,III Bowen lines ($\lambda=5007$~\AA, top left, and $\lambda=4559$~\AA, bottom right)
in the spectrum of ${\rm NGC}\,2392$. The faint horizontal bands show the fragments of the continuum spectrum
of the central star. The spectra of image details projected on different areas along the height of the NES slit are
can be seen along the vertical axis and the splitting (fragmentation) of lines due to line-of-sight velocity differences,
along the horizontal line.}
    \label{NGC2392}
\end{figure*}

The study of non-synchronized intensity variations in various
Balmer lines allows three effects to be separated: suppression of
emissions by molecular absorption in the envelope, suppression in
metal absorption lines, and variations of the emission intensities
proper. Note that the half-width of undistorted emissions (more
than 1\,\AA) is several times greater than both the thermal
Doppler broadening and the broadening arising in the emission
region behind the spherically symmetric shock. Thus, the broad
simultaneously recorded fragment of the spectrum and the high
spectroscopic resolution of the NES spectrograph also open up the
possibility of studying asymmetric mass loss in AGB-stage stars.

\section{YOUNG PLANETARY NEBULAS}

The rapid transition from the red giant   to the planetary
nebula (PN) stage \citep{Harm}, the transition
duration is shorter than $10^4$~years) makes it difficult to find
and study the rare  post-AGB objects. \citet{Kwok} suggested (and
this hypothesis was later confirmed) that some signs of mass loss
in red supergiants may still remain apparent during the young (compact) PN
stage. Remnants of matter lost at the AGB and post-AGB stages have
masses comparable to those of the PN, and are an important factor
in the formation of the observed PN structure. Studies of the
kinematic structure of PNs, like spectroscopy of Mira envelopes,
will help clarify the picture of mass loss during the fast phase
of the crossing the gap of the H--R diagram by low-mass luminous
stars. Here, we consider our primary goal to be to improve the
accuracy of determination of the kinematic expansion velocities of
PN fragments observed in the lines of different elements and their
ions, followed by obtaining (a lower) estimate for the dynamic age
of the nebula. This can bring about the main advantage of the NES
spectrograph over most spectrographs used for producing catalogs
of PN expansion velocities \citep{Sabbadin, Weinberger}.
Figure\,\ref{Vy-Oxy} shows the O\,III Bowen line profiles in the spectrum
of ${\rm Vy}\,1$-$1$ obtained with the NES spectrograph. The
OIII\,$\lambda\,4559$~\AA\ line profile is marked in red, and the
OIII\,$\lambda\,5007$~\AA\ profile,  in black. Half the
distance between the two dashed verticals in this figure
corresponds to the expansion velocity.  The accuracy of the
Doppler component difference $\Delta V_r$ is $150$~m/s.

The advantage of the NES echelle spectrograph is that it allowed
observations to be made in the ``high slit'' mode, which makes it
possible to compare the expansion velocities measured from the
lines of different atoms and ions for different position angles of
the slit position in the PN image and different points along the
height of the slit. For illustration, Figure~\ref{NGC2392} shows a
Doppler image of a pair of O\,III Bowen lines ($\lambda$=5007\,\AA\,
(top left) and $\lambda$=4559\,\AA\,(bottom right)) in
the spectrum of NGC\,2392 -- a natural laboratory of jet
formation. The spectra of image details projected on different
areas along the height of the NES slit are can be seen along the
vertical axis and the splitting (fragmentation) of lines due to
line-of-sight velocity differences, along the horizontal line.

A combinatiion the NES spectrograph  with a 4.6K$\times$2K
detector allows  the PN spectrum to be recorded to register (in two exposures)
in the wavelength interval spanning from the ground-based ultraviolet to the
near infrared, making it possible to study the lines formed by combinations of
different processes (excitation, ionization, collisions, recombination, and fluorescence).

\section{CONCLUSIONS}

The design features and parameters of the NES spectrograph (fixed
location in the telescope, high spectroscopic resolution, wide
spectral range, the possibility of observing extended objects)
serve both as the foundation for performing a variety of
spectroscopic studies, and a stimulus for further development of
the spectrograph, including the first steps along the lines of
adaptive optics \citep{Adapt}. The most productive programs proved
to be those combining elements of the survey with long-term
monitoring.

\section*{Acknowledges}
The work of  V.G.~Klochkova was supported by the Russian Science Foundation (project No.~20-19-00597).
V.E.~Panchuk acknowledges the support from Government of the Russian Federation and the Ministry of
Higher Education and Science of the
Russian Federation (grant~075-15-2020-780) (No.~13.1902.21.0039). Observations on the telescopes
of the Special Astrophysical Observatory of the Russian Academy of Sciences
are supported by the Ministry of Science and Higher Education of the
Russian Federation (contract No.~05.619.21.0016, unique project identifier
RFMEFI61919X0016).

\end{document}